# Relative Positions of Countries in the World of Science

Seyyed Mehdi Hosseini Jenab[1] & Ammar Nejati[2]

[1]*Physics Department, Amirkabir University of Technology, P.O. Box 15875-4413, Tehran, Iran*

*jenab@aut.ac.ir*

[2]*Physics Department, Sharif University of Technology, P.O. Box 11365-9161, Tehran, Iran*

*'These authors contributed equally to this work'*

## Abstract

**A novel picture of the relative positions of countries in the world of science is offered through application of a two-dimensional mapping method which is based on quantity and quality indicators of the scientific production as peer-reviewed articles. To obtain such indicators, different influential effects such as the background global trends, temporal fluctuations, disciplinary characteristics, and mainly, the effect of countries resources have been taken into account. Fifty countries with the highest scientific production are studied in twelve years (1996–2007). A common clustering algorithm is used to detect groups of co-evolving countries in the two-dimensional map, and thereby countries are classified into four major clusters based on their relative positions in the two-dimensional map. The final results are in contrast with common views on relative positions of countries in the world of science, as demonstrated by considering some examples like USA, China or New Zealand. The proposed method and results thereof might influence the concept of 'scientific advancement' and the future scientific orientations of countries.**

Quantitative analysis of the productions of research units (countries, academic institutions, or research groups) has turned out to be one of the most significant issues since the early days of scientometrics. An important part of that analysis is devising methods and indicators to evaluate the multitude of aspects of such a complex production process, especially, its 'quantity' and 'quality' [*e.g.* 1, 2]. This quantitative assessment is becoming inevitable today, since such methods and findings are invaluable for policy-makers of research units as well as academic agents directly involved in the process, to evaluate their achievements, and modify their approaches for further progress. As a consequence, a vast amount of research has been devoted to the development of such methods and indicators, and their critiques.

One of the essential problems in such analyses is that to provide a holistic perspective from scores of different indicators – each representing a certain aspect of the scientific output – usually, a host of graphs and tables are offered, from which one can hardly grasp a comprehensive picture and an ultimate judgment [*e.g.* 3, 4, 5, 6]. As a way out of this problem, composite indicators [7] and two-dimensional maps have been introduced [8, 9, 10].

Along the same path, in this article, we have propounded a method to construct a 2-dimensional map from two composite indicators – the first based on publications and the second based on citations thereof – aiming at a simultaneous representation and analysis of 'quantity' and 'quality' of the scientific productions of countries. This 2-dimensional map is meant to indicate the relative positions of countries compared to each other. However, certain adjustments should be made in advance to take into account influential effects such as the background global or discipline-characteristic trends as well as temporal fluctuations, and mainly, the effect of different amount of resources accessible to countries. Regarding this, we have recognised the effect of human and financial resources upon those relative positions, and have shown that their elimination will provide utterly different positionings which are in sharp contrast with the common views reflected in the national or international reports on scientific production of countries [*e.g.* 3, 4, 5, 6].

In this study, the topmost 50 countries in scientific production are considered in a period of twelve years (1996–2007). Since this period will certainly admit temporal fluctuations in positions, we have implemented a common statistical clustering method (*k-means* algorithm) to analyse the 2-dimentional map (which is unbiased by the resource effects). The finest clustering scheme has been selected by inspection of clustering validity indices. The final clusters (labelled by letters A, B, C, and D) – the objective of this study – indicate the relative positions of country groups compared to each other. We have deliberately avoided the common convention of proposing some clear-cut numbers as 'ranks' or 'position markers' for countries, mainly because there is actually an insignificant difference between countries belonging to a neighbourhood in the range of values of an indicator; moreover, those numeric ranks prove to be volatile due to inevitable and non-negligible temporal fluctuations.

## I. Method

The proposed method involves a modified version of the previous scientometric indices to measure the quality and quantity of the *explicit* scientific production of a research unit in form of journal papers. In most of the scientometric studies, indices based on number of papers and their citations are used as measures of quantity and quality of scientific production, respectively. By acquiring such indicators, one can obtain a graphic representation of the data by plotting the quantity indicator on the horizontal axis, and the quality indicator on the vertical axis to produce a two-dimensional 'quantity–quality map'. However, before coming to that point, certain subtleties ought be taken into account.

**(I1) Effect of Resources.** Countries have widely different financial and human resources which have definitely a strong influence on the quantity and quality of their scientific production, and therefore, on their relative positions in the final quantity–quality map. As the first step, one can construct a map without considering the different resource sizes accessible to countries (see §II1). Quite obviously, a country with large resources (like USA or China) is likely to publish more, and obtain more citations than a country with much smaller resources (like New Zealand). However, our main objective here is to analyse scientific performance of countries while taking into account the share of total resources they benefit. Therefore, to achieve an equitable picture of the relative positions of countries in the scientific world – which would be unbiased by the size of resources – we have to account for these resource effects in the quantity and quality indicators. However, it appears that the resources affect the quantity and quality indicators in different manners.

*(I1a) Publication*. In a paradigmatic society, an increase in resources would be doubtlessly accompanied by a growth in the absolute number of publications. In fact, we have supposed that a typical country should contribute a share of scientific publications directly proportional to its share of total resources of the countries considered in the study. To account for the effect of resources, firstly, one has to combine the human and financial resource measures into a single composite resource indicator; a straightforward way is to render them dimensionless and subsequently sum them up into a resource indicator, *R*:

$$R_\nu(t) = \frac{Pop_\nu(t)}{\sum_{\nu=1}^{N} Pop_\nu(t) \Big/ N} + \frac{GDP_\nu(t)}{\sum_{\nu=1}^{N} GDP_\nu(t) \Big/ N} \qquad (1),$$

where $N$ is the number of countries (labelled by $v$) and *Pop* stands for population. Here, total population and GDP (gross domestic product) have been used, respectively, as indicators for the human and financial resources of countries. Division by the average taken over all the studied countries in the same year, is used in converting them to dimensionless numbers.

By dividing the publication number (as a quantity indicator) by the resource indicator, a new quantity measure, publication per resource (PPR), will be obtained:

$$PPR_v(t) = \frac{P_v(t)}{R_v(t)} \qquad (2).$$

*(I1b) Citation*. The influence of resources on citations is rather indirect, since in a paradigmatic society increasing the resources would firstly beget a rise in ‘quantity’ and this would, in turn, lead to more citations. However, a rise in citations does not mean an increase in overall quality because a high number of citations might indicate either a high quality of scientific production, or a high number of low quality productions. Hence, to obtain an honest measure of quality, one has to remove the second effect which might be a consequence of an increase in resources. For this purpose, we have utilized the well-known indicator, citation per publication (CPP). In this study, the extent of the ‘citation window’ is from the year of publication ($t$) to the final year of the studied period ($t_f$, here 2007). CPP of a country in a specific year is the total citations to publications of a country within the citation window, divided by the total publications of that country in the same year:

$$CPP_v(t) = \frac{C_v(t;t_f)}{P_v(t)} \qquad (3).$$

Regarding the previous discussions on publication and citation (I1a & I1b), to take into account the effect of resources in our quantity and quality measures, we use PPR and CPP, respectively.

**(I2) Effect of Background Global Trends on Publications and Citations.** When considering the indicators in a period of time, the global trends in quality and quantity of scientific production will affect the positions of countries in the maps, and only by their elimination, the positions in different years will be comparable to each other.

*(I2a) Temporal Growth in Publication*. An annual increase in number of publications and in publication per resource (PPR) is observed throughout the world (Fig. 1). Thus, an absolute temporal increase in scientific production of a country does not necessarily mean a rise in quantity *compared to the world*.

*(I2b) Intrinsic Trend of Citation*. It has been empirically observed that one should wait some years after the publication of a paper for the citations to grow considerably. Regarding this fact, a diminishing trend will be seen for the citations of papers and consequently, in citation per publication (CPP) near the end of the study period (here, 2007) (Fig. 2). This intrinsic decreasing trend would conceal and misrepresent the actual temporal evolution of quality of the scientific production of a research unit.

To eliminate these two background trends (I2a & I2b), a scaling method has been applied in which the value of every indicator in a specific year is divided by the average value of that indicator in the same year. The average is taken over all the fifty countries in that year:

$$PPRm_{\nu}(t) = \frac{PPR_{\nu}(t)}{\overline{\overline{PPR}}(t)} \qquad (4a),$$

$$CPPm_{\nu}(t) = \frac{CPP_{\nu}(t)}{\overline{\overline{CPP}}(t)} \qquad (4b),$$

where

$$\overline{Q}_{\nu}(t) = \frac{1}{N}\sum_{\nu=1}^{N} Q_{\nu}(t) \qquad (4c).$$

Here, '*m*' indicates a scaling by the average value. This scaling-by-average routine helps to remove the effects of the background global trends and provides dimensionless numbers that can be used in comparisons across different years and disciplines.

**(I3) Discipline-Specific Characteristics of Different Branches**. When combining or comparing indicators corresponding to different branches of science, one should note that each branch of science has its own distinctive trend of publication/citation. For instance, mathematical branches have a lower publication/citation than medical ones. Therefore, to make a fair comparison of the quality of a part (or whole) of scientific productions of countries, one should doubtlessly take account of this fact, since, *e.g.* a higher absolute publication/citation number in a discipline does not necessitate a better performance relative to another discipline with a lower absolute publication/citation. Here again, the auxiliary scaling method used in §§I2–3 can resolve this issue and thereafter, quantity or quality indicators of different disciplines will be comparable to each other. Hence, to arrive at indicators that can be compared across different branches, initially, we have calculated the scaled indicator *Qm* (here, *Q* might represent any indicator like PPR or CPP) for every branch in each year; then, we have averaged over the *Qm*'s corresponding to different branches which make up a discipline (see §I4). In this study, we have used the *Scopus* classification of science into 27 branches [11]. Therefore, *e.g.* when considering the whole body of science, all the 27 branches are averaged over:

$$Qa_{\nu}^{\text{discipline}}(t) = \frac{Qm_{\nu}^{\text{branch}\ \alpha}(t) + Qm_{\nu}^{\text{branch}\ \beta}(t) + \cdots}{\text{total number of branches involved in } \mathrm{Q}^{\text{discipline}}} \quad (5);$$

for example,

$$PPRa_{\nu}^{whole\,science}(t) = \frac{PPRm_{\nu}^{physics}(t) + PPRm_{\nu}^{mathematics}(t) + \cdots}{\text{total number of branches in whole science (27)}} \quad (6a),$$

$$CPPa_{\nu}^{whole\,science}(t) = \frac{CPPm_{\nu}^{physics}(t) + CPPm_{\nu}^{mathematics}(t) + \cdots}{\text{total number of branches in whole science (27)}} \quad (6b).$$

**(I4) Clustering Method.** The aforementioned procedure (§§I1–I3) provides a time-series of PPR*a*-CPP*a* data for countries throughout the studied period. Here, we will have 600 points on the 2-dimensional map constructed from PPR*a*-CPP*a* data of 50 countries during 12 years (1996–2007) for each branch or discipline of science. However, this map is still too intricate to decipher, mainly due to the complexities of temporal evolution of countries. Therefore, to obtain an intelligible picture out of this map, we have used a clustering analysis to find out whether some groups of co-evolving countries can be detected.

From the plenty of clustering algorithms available, we have selected and applied the well-known *k-means* algorithm. K-means is a partitional clustering algorithm widely used in a variety of applications, and it is relatively easy to understand and implement [12]. However, clustering algorithms, like k-means, are unsupervised processes with no predefined classes or prototypes, and the final number of clusters is not given within them. Hence, we have used two internal validity indices (*Dunn's* and *SD* criteria) to find out the optimum number of clusters. These criteria measure the compactness and separation of clusters; moreover, they evaluate the clustering results using only quantities and features acquired from merely the data set [13]. The optimum number of clusters would be the one corresponding to the lowest indices.

**I5. Data.**

*(I5a) Data Sources*. The data needed for the analyses have been obtained from the *SCImago Journal & Country Rank* portal [14] which provides Scopus data arranged according to country, branch of science and year. The population and gross domestic product (GDP) data have been obtained from the *World Development Indicators* database [15].

*(I5b) Categorization of Branches*. We have categorized the initial 27 branches of science into 6 major disciplines: (1) Natural Sciences (*Nat*): Physics and Astronomy, Neuroscience, Immunology and Microbiology, Earth and Planetary Sciences, Chemistry, Biochemistry, Genetics and Molecular Biology, Agricultural and Biological Sciences; (2) Formal Sciences (*Frm*): Mathematics, Computer Science; (3) Engineering Sciences (*Eng*): Materials Science, Environmental Science, Engineering, Energy, Chemical Engineering; (4) Health Sciences (*Hlt*): Veterinary, Pharmacology, Toxicology and Pharmaceutics, Nursing, Medicine, Health Professions, Dentistry; (5) Social Sciences (*Soc*): Decision Sciences, Social Sciences, Psychology, Economics, Econometrics and Finance, Business, Management and Accounting; (6) Arts and Humanities (*Arh*): Arts and Humanities. Indeed, this categorization is arbitrary.

## II. Results and Discussion

**(II1) Map *with* the Resource Effects.** As a primary step and for the sake of comparison, we have presented a map in which we have removed the effect of the background global and discipline-characteristic trends on the quantity and quality indicators by the scaling-to-average method of §§I2–3; *i.e.* influence of resources has not been accounted for. Two indicators are used to construct the map, namely, P*a* as the quantity and C*a* as the quality indicator:

$$Pa_{\nu}(t) = \frac{\sum_{\{branches\}} Pm_{\nu}^{branch}(t)}{27} \quad \text{(7a)},$$

$$Ca_{\nu}(t) = \frac{\sum_{\{branches\}} Cm_{\nu}^{branch}(t;t_f)}{27} \quad \text{(7b)}.$$

Plotting these indicators on a 2-dimensional map (P*a*–C*a* map), we obtain a rather revealing picture. Huge gaps were evident between groups of collocated countries so that we could recognise 3 groups according to the positions of countries. Fig. 3 shows the whole picture by three figures of different scale. In this figure we have shown the temporal average of P*a* and C*a* for each country to avoid the complexity due to temporal fluctuations. The detailed information about the relative positions of countries in this map are presented in the last column of Table 1.

**(II2) Map *without* the Resource Effects.**

*(II2a) The Map.* In the second step, we have accounted for the effects of the resources by using PPR and CPP (as described in §I1) as our quantity and quality indicators. Furthermore, we have applied the scaling method (§§I2–3) to remove the effect of global and discipline-specific trends on these indicators (hence, PPR*a* and CPP*a*) from which a 2-dimensional map has been constructed. This time, a totally new picture is exposed which is much more intricate than the previous P*a*–C*a* map.

At the final step, we have applied the clustering method (§I4) to this map to extract the inherent geometrical structure which would reveal the country clusters. The clustering validity indices indicated that the best clustering result is composed of 4 clusters which we have labelled by letters A, B, C and D (to be discussed below).

Our focus of study is the whole body of science; yet, to offer a more detailed picture which includes also a rough measure of the relative positions of countries in different disciplines, we have applied our method to the 6 disciplines of science (§I5b). It is remarkable to note that the validated clusterings for the disciplines reveal almost the same pattern as that for the whole science; *i.e.* four clusters with the same relative positioning. This is especially evident in natural, health and engineering disciplines. In the case of other disciplines (formal and social sciences, and arts and humanities) the clusters are not as distinct and separated as in the previous. Nevertheless, the cluster labels in all disciplines have the same meaning (as regards the position in the map) as those for ‘Whole Science’; for example, in social sciences (‘SOC’) there is no cluster C, because in that case no cluster is located near cluster C of ‘Whole Science’, but two clusters (of social science) are located near cluster A of ‘Whole Science’ (Fig. 5) and hence, are labelled as A.

*(II2b) Clusters.* The four clusters A to D are representatives of the global quantity–quality classes (Fig. 4). Members of cluster A are in an excellent level of both quantity and quality. Cluster B includes countries with rather the same quality as cluster A, but lower quantity. Cluster C is composed of countries with almost the same quantity level as that of cluster B, but lower quality level; thus, cluster C has a lower level in both quality and quantity in comparison with cluster A. Members of cluster D are in the lowest level in both quality and quantity compared to all the other three clusters. Some countries are ‘transitional’ which means that they move from one cluster to another in the course of their temporal evolution, and thence, are marked by two (or more) cluster labels.

Table 1 shows the results of clustering for the whole body of science together with those for the 6 major disciplines. Countries are arranged according to, in the first place, their cluster labels in whole science (shown under ‘Whole Science’), and in the second, their cluster labels in the major disciplines. It should be reminded that the position of a country within the ‘Whole Science’ cluster is arbitrary to some extent, and they shall not be considered as a rigid ranking as discussed in the Introduction. This arbitrariness is, firstly, due to the fact that sorting the countries based on other preferences on the ordering of disciplines will alter the positions, and secondly, that other definitions for disciplines (§I5b) can alter the positions within the ‘Whole Science’ clusters.

*(II2c) Comparison with Pa-Ca Map.* Fig. 4 shows the relative positions of countries in the map without the resource effects. Compared to the Fig. 3 (P*a*-C*a* map) which includes that effect, we can notice that the huge gap between countries has disappeared; moreover, positions of some countries have been changed drastically.

One of the countries which undergoes a drastic change in position upon the elimination of resource effects is USA. In Fig. 3, it has a superior position relative to all others, especially its European rivals like UK and Germany. However, in Fig. 4, it appears as a group-B member with a considerable distance from superior countries like Israel, Sweden or New Zealand. This is a remarkable instance of the consequences of elimination of the resource effects, in which countries with lower resources outperform a country like USA which has a considerable share of the global scientific production. Another significant instance is the case of China. It has a superior position (especially, as regards publication number) in Fig. 3. Contrastingly, in Fig. 4, it has been demoted to a very low position within group D, with a large distance from the average position in the world of science. A different instance is the case of New Zealand. In Fig. 3, it has a very low position in sharp contrast to its top position in Fig. 4 where it is placed among the excellent members of group A.

In summary, it can be observed that accounting for the effect of resources drastically alters the relative positions of countries as the largest research units, and reveals some previously-unseen features, namely, co-evolving clusters of countries which are observable in the whole body as well as disciplines of science. This viewpoint could radically undermine the current convictions on the positioning of countries in the world of science and affect notions like 'scientific advancement' and its relevant measures. Finally, this perspective might, quite optimistically, foster some positive modifications in the current science policies of countries and their future directions.

Acknowledgements. The authors would like to express their thanks to A. Bavali, R. Ghasemi, F. Jahangiri, and Z. Rezaee for the fruitful discussions in the preliminary stages of the study which led to this article.

**Table 1 Clustering results for PPR*a*-CPP*a* map (which is unbiased by the resource effects) for the 'Whole Science' and its 6 major disciplines (see §I5b). The last column shows the relative position of countries in the P*a*-C*a* map (which is biased by the resource effects). Some countries are 'transitional', and thence, are marked by two (or more) cluster labels.**

| Name | Code | PPR*a*–CPP*a* Map | | | | | | | P*a*–C*a* Map |
|---|---|---|---|---|---|---|---|---|---|
| | | ***Whole Science*** | Science Disciplines | | | | | | |
| | | | *Nat* | *Eng* | *Hlt* | *Soc* | *Frm* | *Arh* | |
| **Israel** | IL | ***A*** | A | A | A | A | A | A | Mi |
| **Sweden** | SE | ***A*** | A | A | A | B | B | A | Mi |
| **Finland** | FI | ***A*** | A | A | A | B | C-B | A | Mi |
| **Australia** | AU | ***A*** | A | B | A | A | B | A | Mi |
| **New Zealand** | NZ | ***A*** | A | B | A | A | C | A | Lo |
| **Canada** | CA | ***A*** | A | B | A | B | C | A | Hi |
| **Switzerland** | CH | ***A*** | A | B | A | B-A | B | B | Mi |
| **UK** | UK | ***A*** | A-B | B | A | A | B | A | Hi |
| **Singapore** | SG | ***A*** | B | A | C-B | B | A | B | Lo |
| **Netherlands** | NL | ***A*** | B | B | A | B | B | A | Mi |
| **Denmark** | DK | ***B*** | A | B | A | B | B | A | Mi |
| **Hong Kong** | HK | ***B*** | B | A | B | A | C | A | Lo |
| **Austria** | AT | ***B*** | B | B | A | A | B | B | Mi |
| **Belgium** | BE | ***B*** | B | B | A | B | B | A | Mi |
| **Germany** | DE | ***B*** | B | B | B | A | B | B | Hi |
| **France** | FR | ***B*** | B | B | B | A | B | B | Hi |
| **Italy** | IT | ***B*** | B | B | B | A | B | B | Hi |
| **Spain** | ES | ***B*** | B | B | B | A | B | B-D | Mi |
| **Portugal** | PT | ***B*** | B | B | B | A | B | D | Lo |
| **USA** | US | ***B*** | B | B | B | B | B | A | Hi |
| **Ireland** | IE | ***B*** | B | B | B | B | B | A | Lo |
| **Norway** | NO | ***B*** | B | B | B | B | B | A-B | Lo |
| **Greece** | GR | ***B*** | B | B | C | A | C | A-B-D | Lo |
| **Korea** | KR | ***B*** | D | C | B | A | B-D | D | Mi |
| **Taiwan** | TW | ***B*** | D | C | B | A | C | B-D | Mi |
| **Slovenia** | SI | ***C*** | C | A | C | A | A | D | Lo |
| **Poland** | PL | ***C*** | C | C | B | D | C | D | Mi |
| **Czech Republic** | CZ | ***C*** | C | C | C | A | C | B-D | Lo |
| **Hungary** | HU | ***C*** | C | C | C | A | C | B-D | Lo |
| **Slovakia** | SK | ***C*** | C | C | C | A | C | D | Lo |
| **Croatia** | HR | ***C*** | C | C | C | A | D | D | Lo |
| **Bulgaria** | BG | ***D*** | C | C | C | D | C-D | D | Lo |
| **Russia** | RU | ***D*** | C | C | D | D | D | D | Mi |
| **Lithuania** | LT | ***D*** | D | C | D | A-D | C | D | Lo |
| **Belarus** | BY | ***D*** | D | C | D | D | D | D | Lo |
| **Romania** | RO | ***D*** | D | C | D | D | D | D | Lo |
| **Ukraine** | UA | ***D*** | D | C | D | D | D | D | Lo |
| **Japan** | JP | ***D*** | D | C-B | B | D | D | B | Hi |
| **Turkey** | TR | ***D*** | D | D | C | D | D | D | Mi |
| **South Africa** | ZA | ***D*** | D | D | D | A | D | B | Lo |
| **China** | CN | ***D*** | D | D | D | D | D | B-D | Hi |
| **Argentina** | AR | ***D*** | D | D | D | D | D | D | Lo |
| **Chile** | CL | ***D*** | D | D | D | D | D | D | Lo |
| **Egypt** | EG | ***D*** | D | D | D | D | D | D | Lo |
| **Iran** | IR | ***D*** | D | D | D | D | D | D | Lo |
| **Mexico** | MX | ***D*** | D | D | D | D | D | D | Lo |
| **Malaysia** | MY | ***D*** | D | D | D | D | D | D | Lo |
| **Pakistan** | PK | ***D*** | D | D | D | D | D | D | Lo |
| **Brazil** | BR | ***D*** | D | D | D | D | D | D | Mi |
| **India** | IN | ***D*** | D | D | D | D | D | D | Mi |

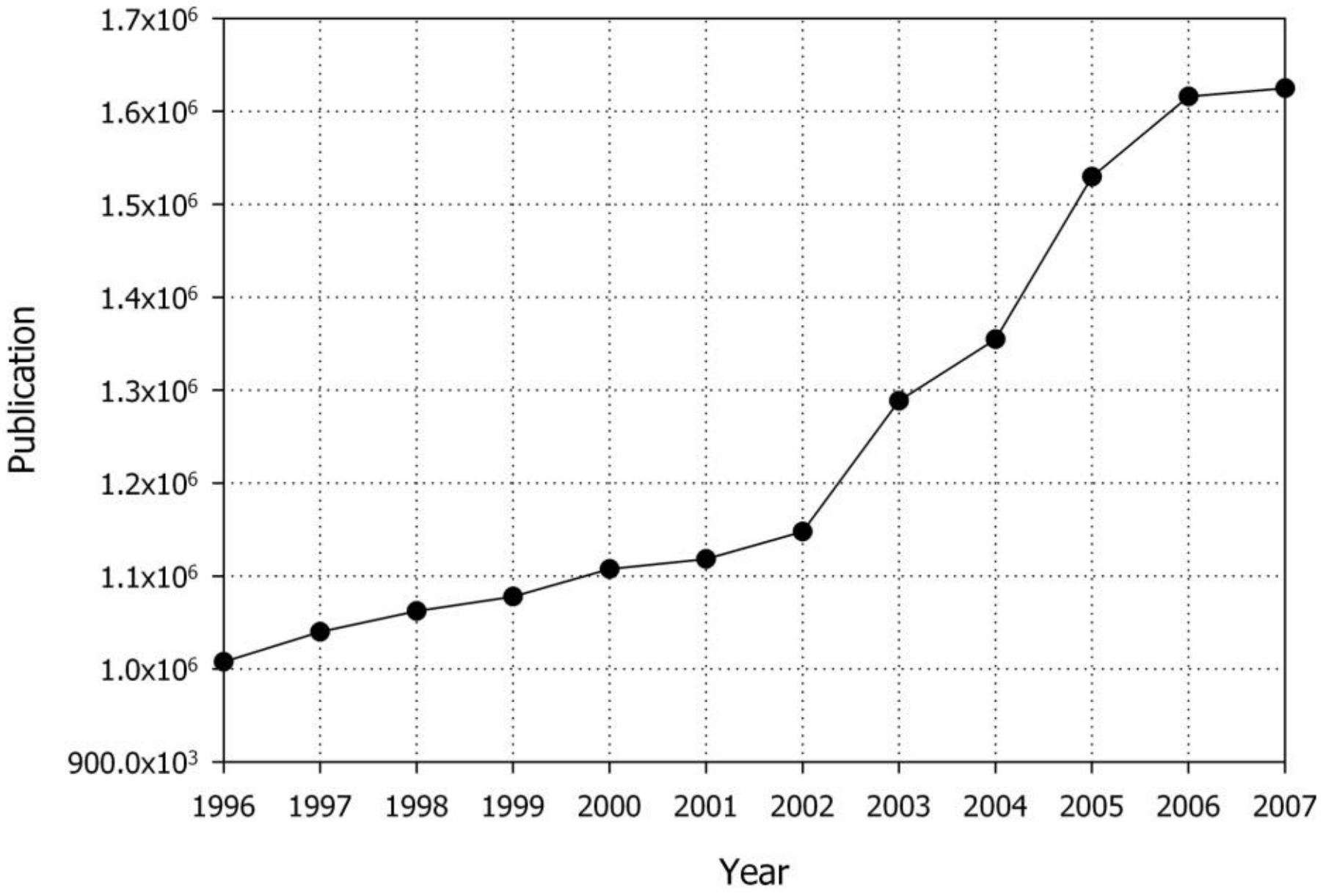

(a)

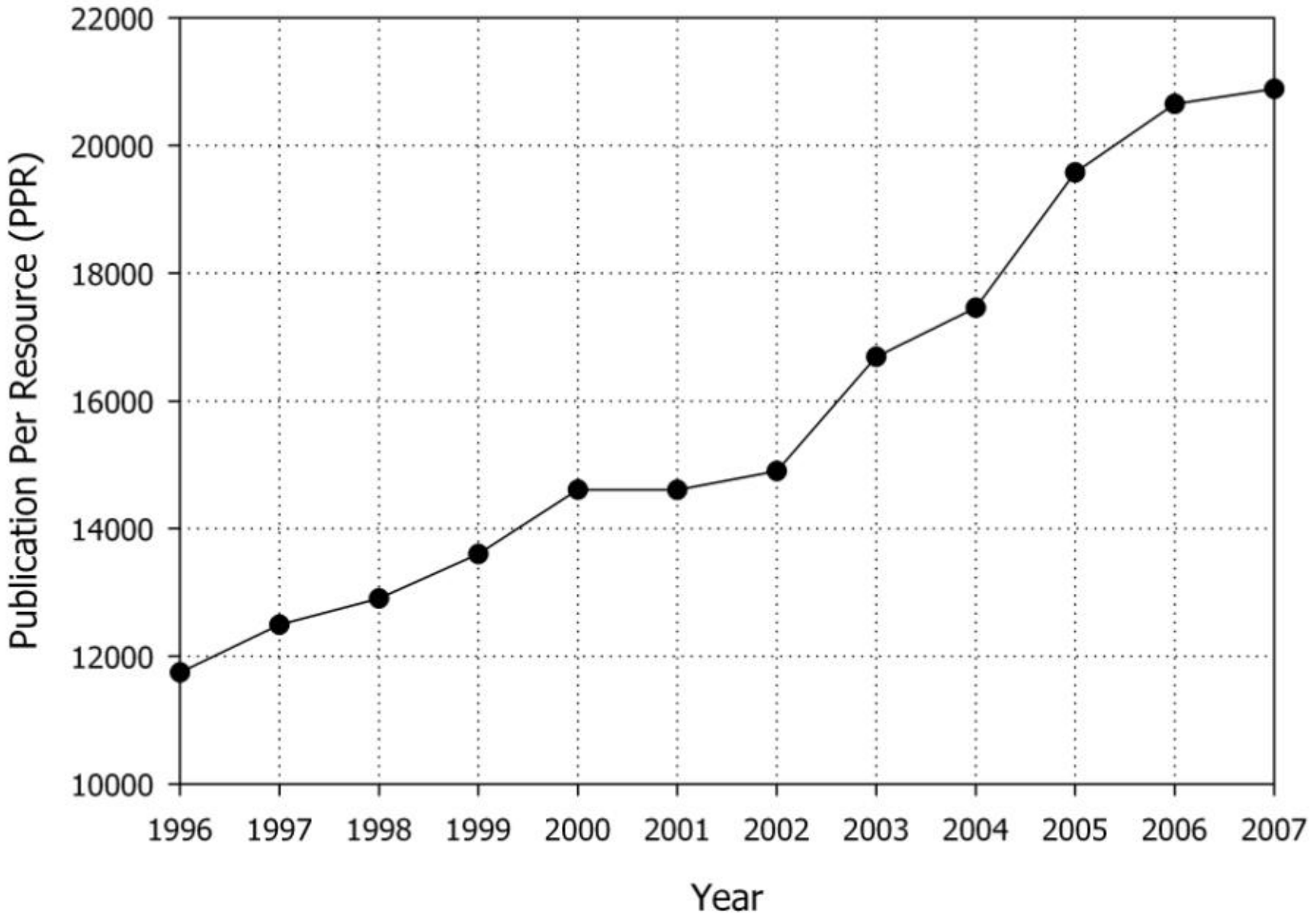

(b)

Figure 1 Temporal growth (a) in publication (P), and (b) in publication per resource (PPR) in the studied period (1996–2007).

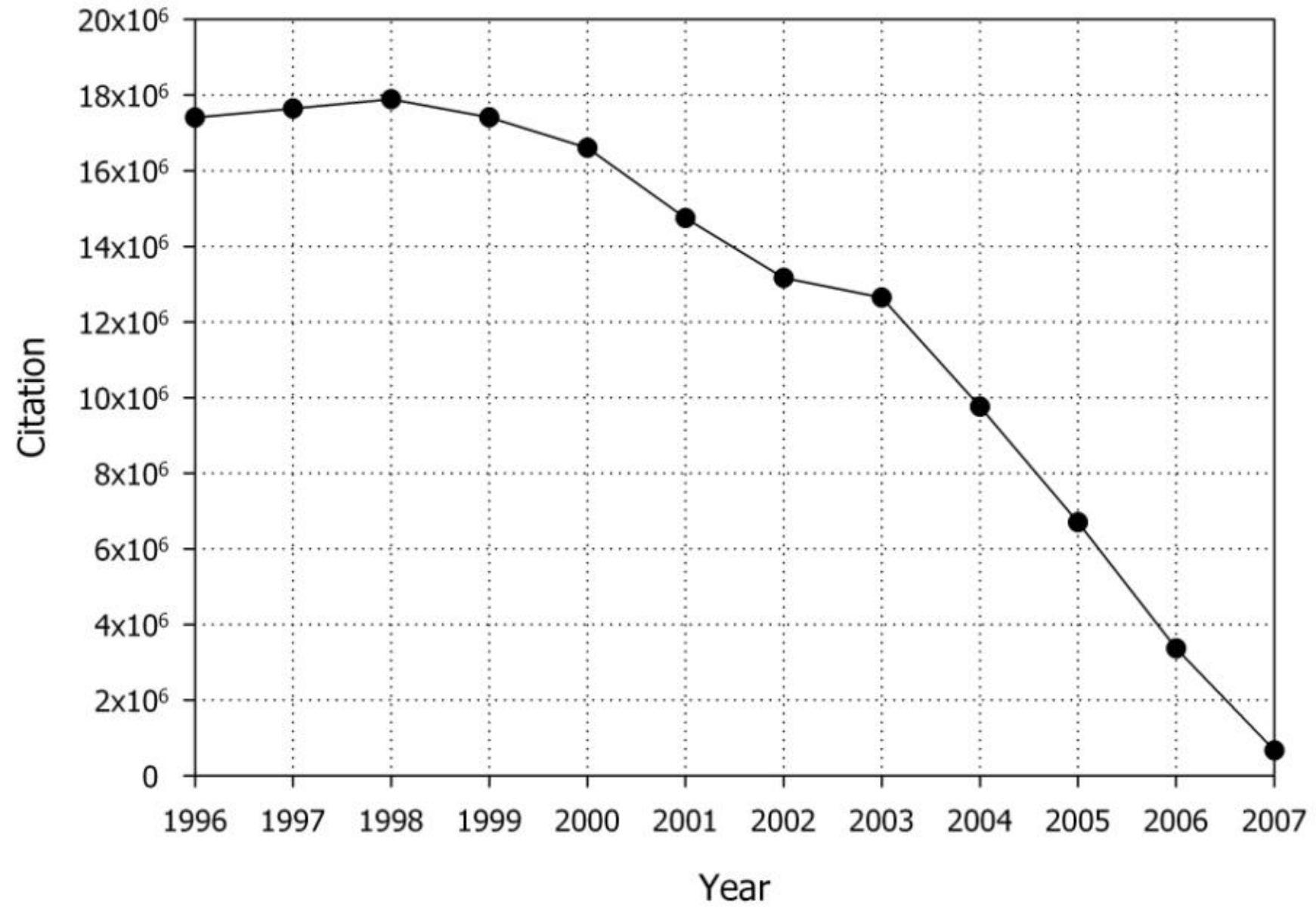

(a)

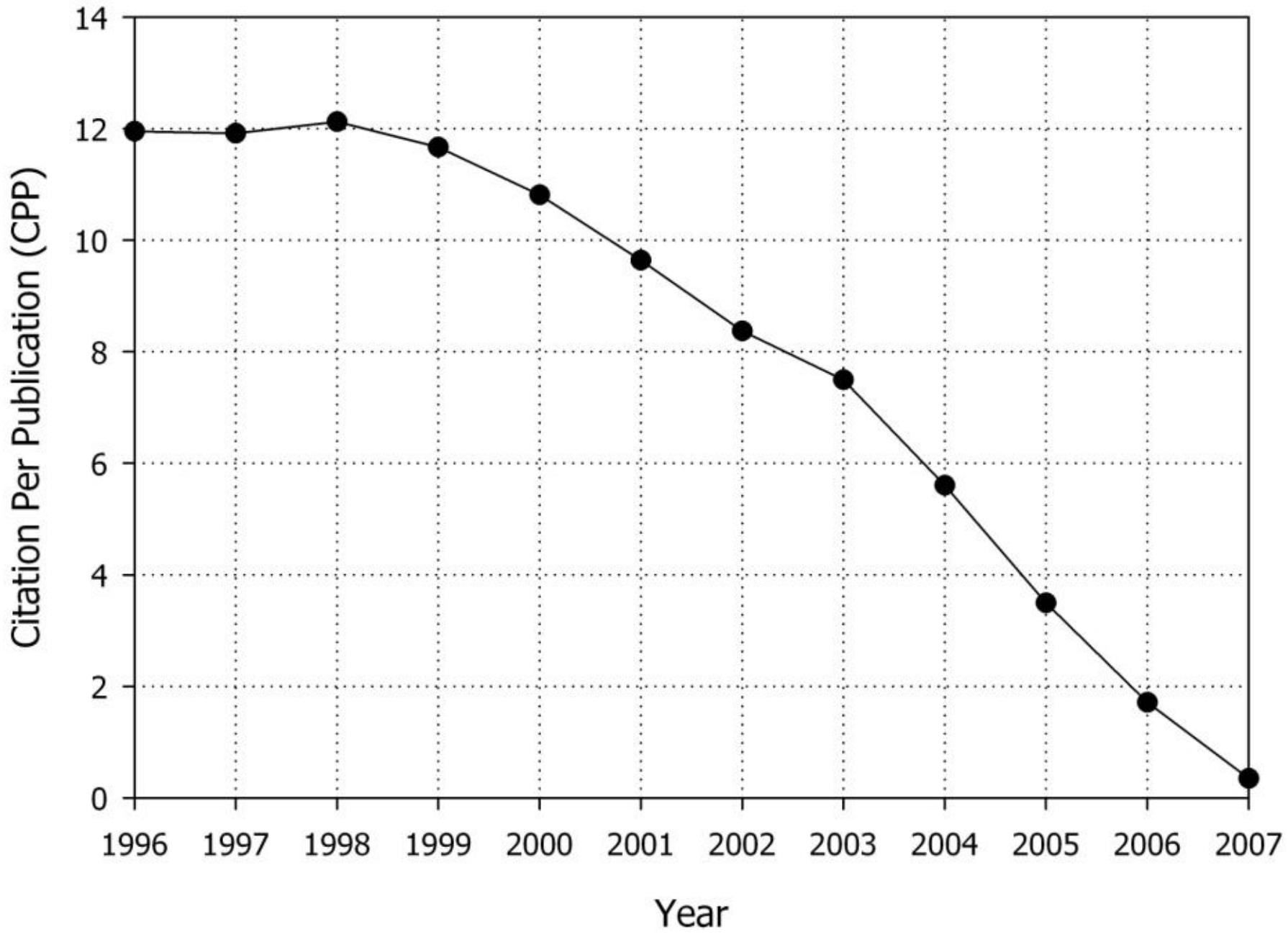

(b)

Figure 2 Temporal decrease in (a) citation, and (b) citation per publication (CPP) in the studied period (1996–2007).

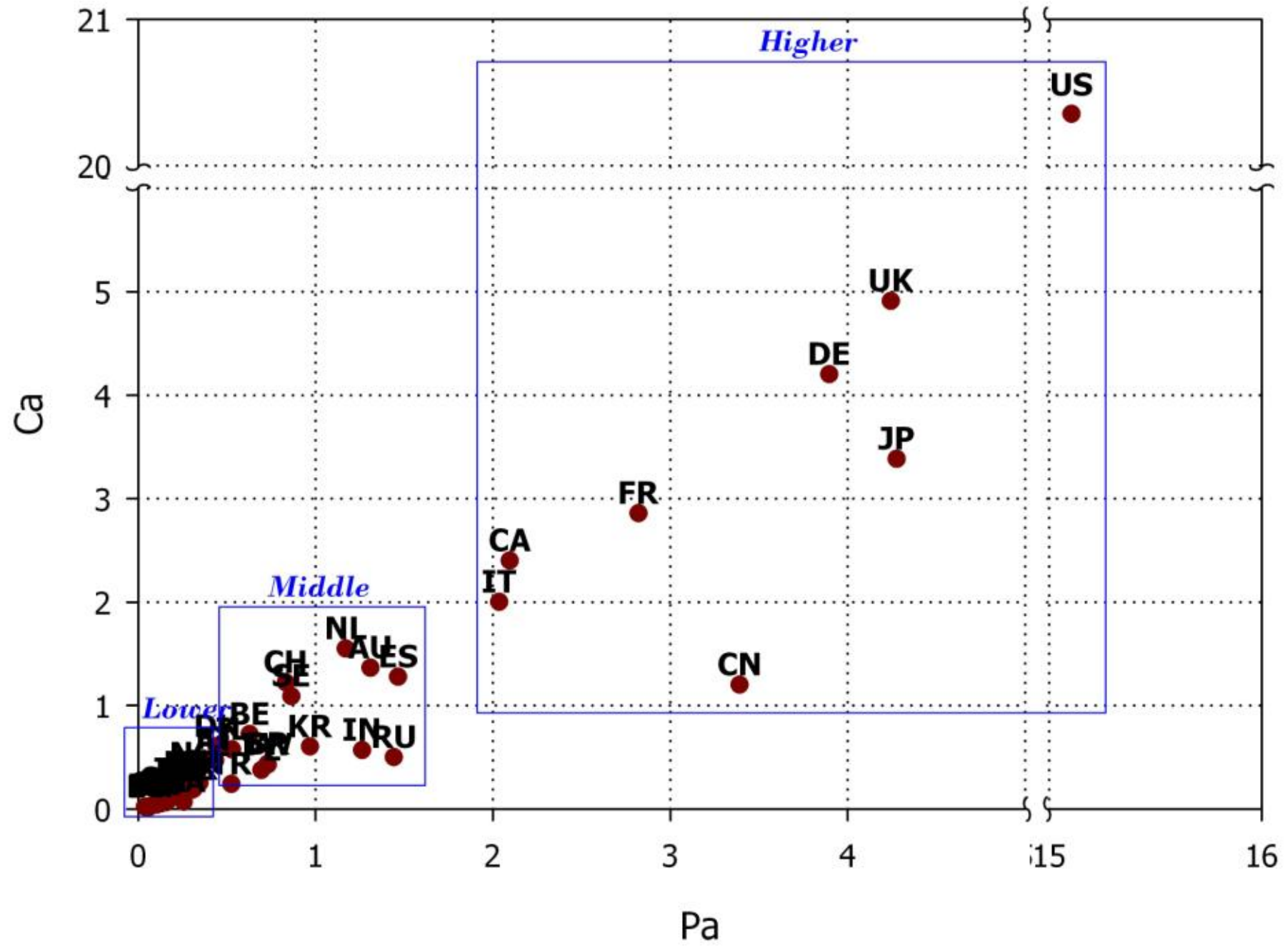

(a)

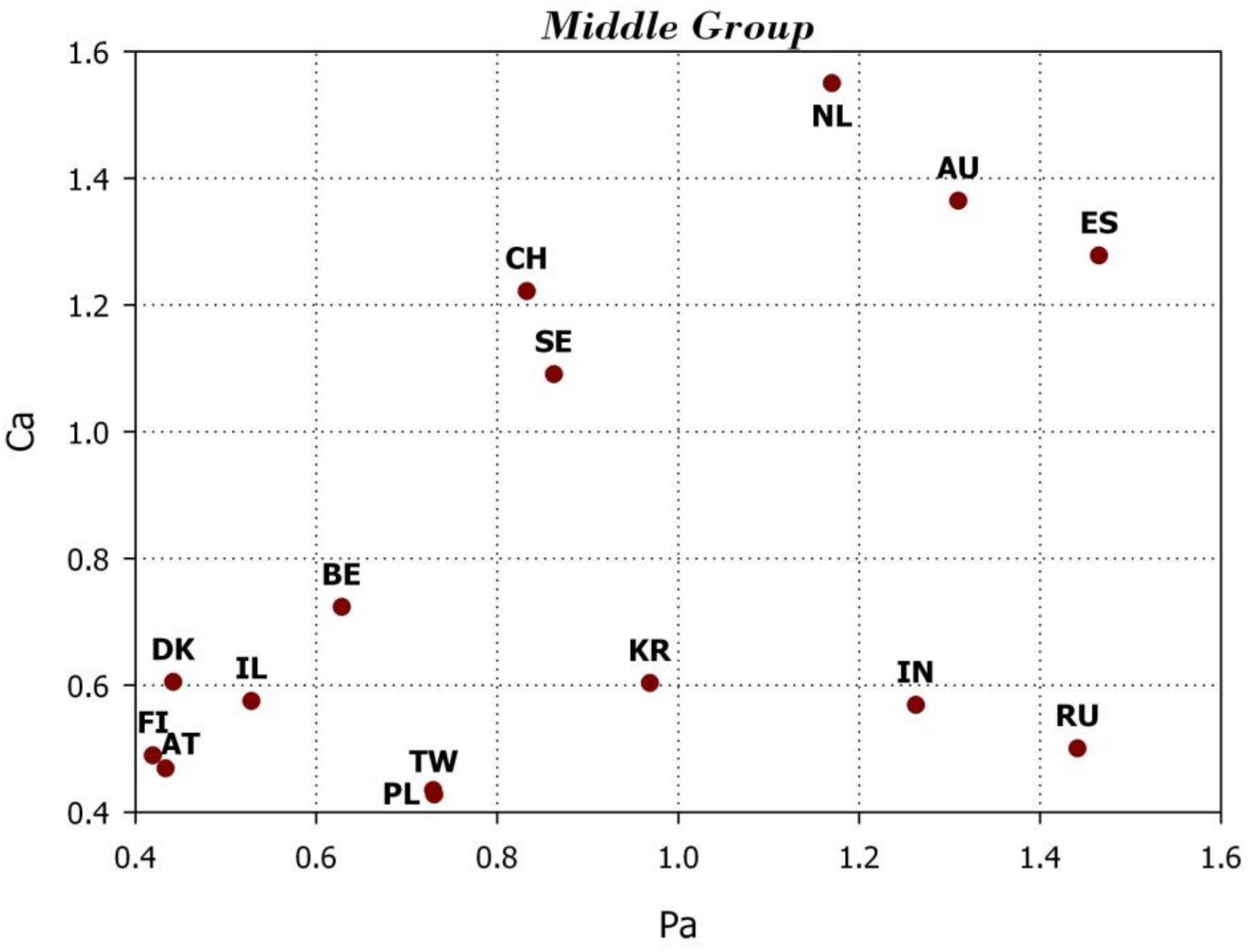

(b)

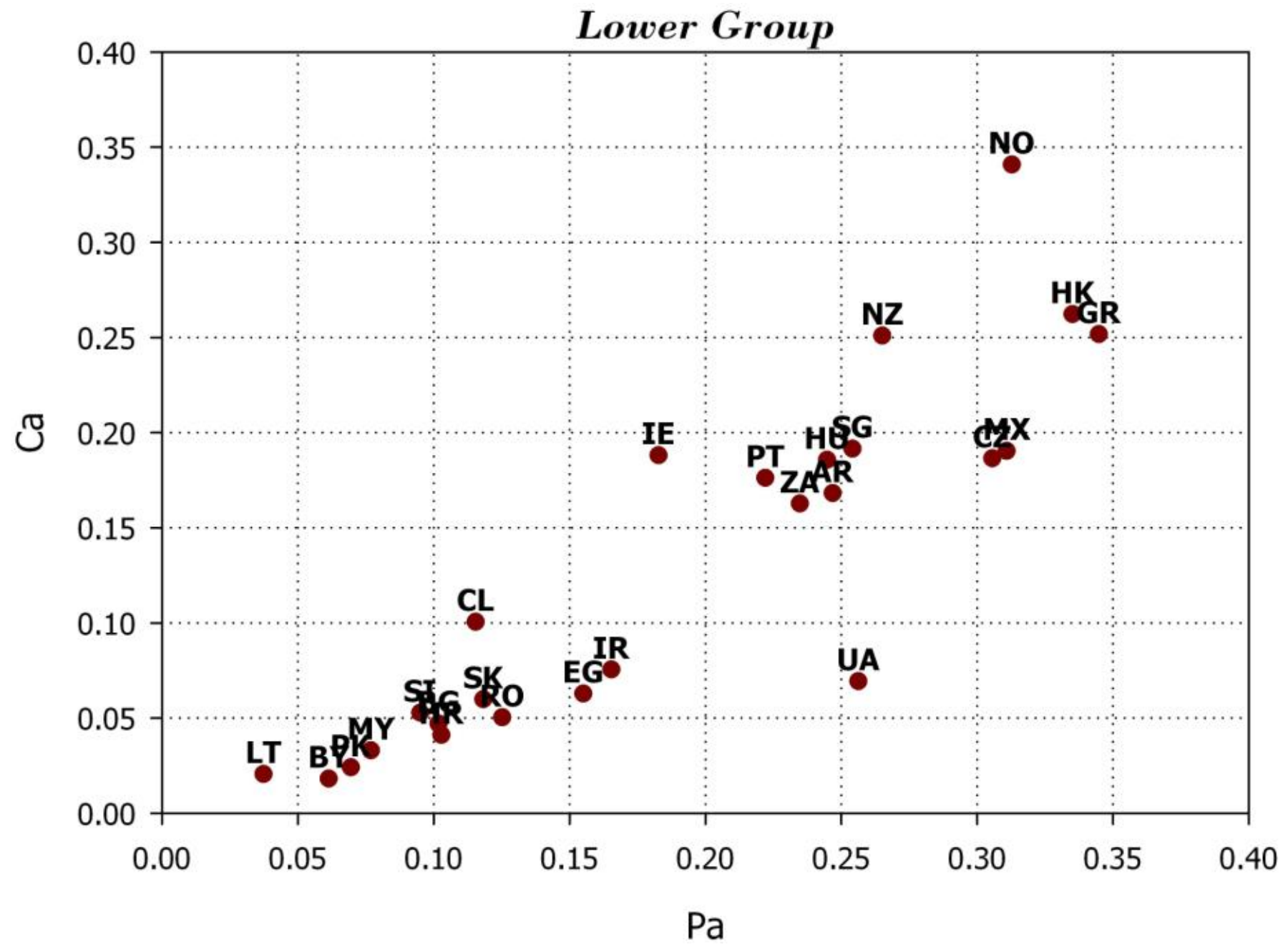

(c)

Figure 3 P*a*-C*a* map (which is biased by the resource effects) for the 50 topmost countries. Each point represents the average value of P*a* and C*a* of a country throughout the studied period (1996–2007). Since the points are scattered in a very wide range for both P*a* and C*a*, the map is presented in three different ranges (3a–3c). (a) three country groups are recognised – "*Higher*", "*Middle*" and "*Lower*" – according to positions in the map. Notice the huge gap between USA and other countries in this map; (b) the "Middle" group; (c) the "*Low*er" group. Refer to Table 1 for further details.

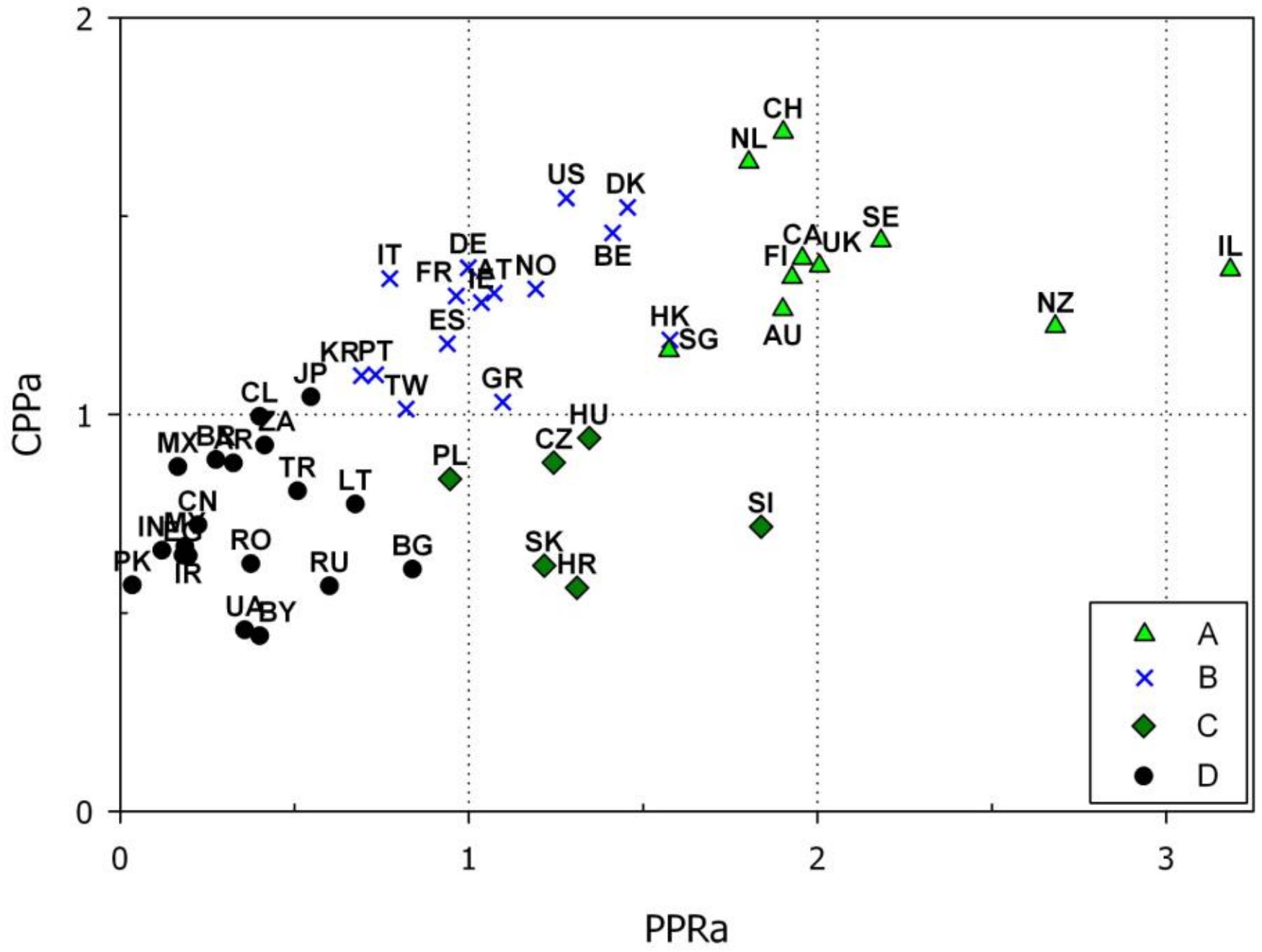

Figure 4 PPR*a*-CPP*a* map (which is unbiased by the resource effects) for the 50 topmost countries. Each point represents the average value of PPR*a* and CPP*a* of a specific country throughout the studied period (1996–2007). Unlike the P*a*-C*a* map (biased by the resource effects) (Fig. 3), there is no huge gap among the countries and relative positions of some countries have changed drastically.

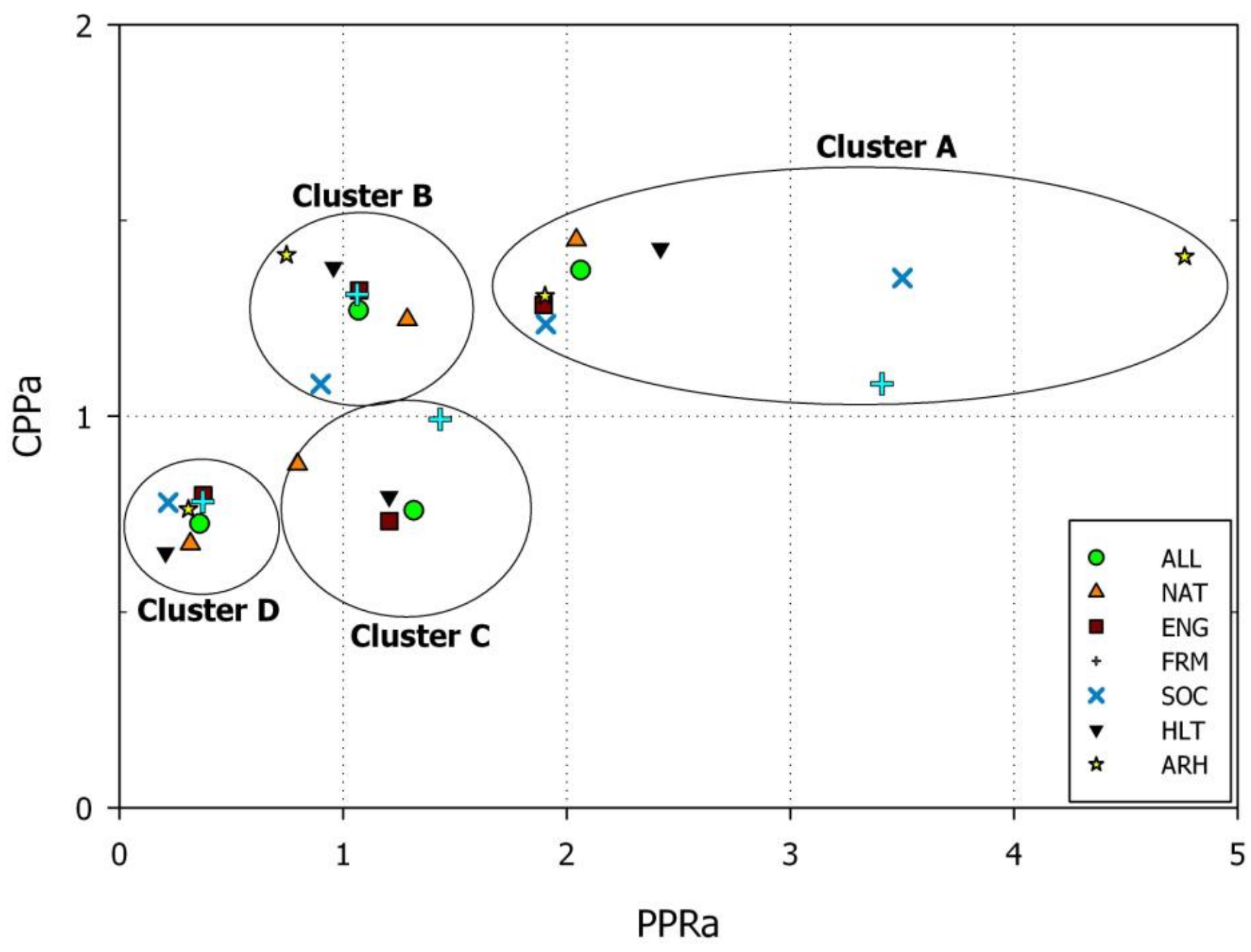

Figure 5 Positions of centroids of four clusters (A–D) for ‘Whole science’ and its 6 disciplines in the PPR*a*-CPP*a* map (Fig. 4).